\documentclass[preprint,eqsecnum,aps,prd,nofootinbib]{revtex4}

\usepackage[dvips]{graphicx}
\usepackage{amsmath}
\usepackage{enumerate}

\def\ben{\begin{equation}}
\def\een{\end{equation}}
\def\bena{\begin{eqnarray}}
\def\eena{\end{eqnarray}}

\newcommand{\I}{{\mathcal I}}
\def\l{\lambda}
\def\a{\alpha}
\def\b{\beta}
\def\be{\begin{equation}}
\def\ee{\end{equation}}
\def\bena{\begin{eqnarray}}
\def\eena{\end{eqnarray}}

\begin{document}
\title {A tale of two superpotentials: \\ Stability and Instability in Designer Gravity}

\author{Aaron J. Amsel${}^{1,}{}$\footnote{\tt amsel@physics.ucsb.edu}, Thomas Hertog${}^{2,}{}$\footnote{\tt thomas.hertog@cern.ch},
Stefan Hollands${}^{3,}{}$\footnote{\tt
hollands@theorie.physik.uni-goettingen.de}, and Donald
Marolf${}^{\,1,}{}$\footnote{\tt marolf@physics.ucsb.edu} }

\affiliation{${}^1$Physics Department, UCSB, Santa Barbara, CA
93106, USA \\
${}^2$Theory Division, CERN, CH-1211 Geneva 23, Switzerland and\\
APC, 10 rue Alice Domon et L\'eonie Duquet, 75205 Paris, France \\
${}^3$Inst. f. Theor. Physik,
Georg-August-Universit\"at, D-37077, G\"ottingen}

\begin{abstract}
We investigate the stability of asymptotically anti-de Sitter
gravity coupled to tachyonic scalar fields with mass at or slightly
above the Breitenlohner-Freedman bound.  The boundary conditions in
these ``designer gravity'' theories are defined in terms of an
arbitrary function $W$.  Previous work had suggested that the energy
in designer gravity is bounded below if i) $W$ has a global minimum
and ii) the scalar potential admits a superpotential $P$.  More
recently, however, certain solutions were found (numerically) to
violate the proposed energy bound.   We resolve the discrepancy by
observing that a given scalar potential can admit two possible
branches of the corresponding superpotential, $P_{\pm}$.  When
there is a $P_-$ branch, we rigorously prove a lower bound on the
energy; the $P_+$ branch alone is not sufficient.  Our numerical investigations
i) confirm this picture, ii) confirm other critical aspects of the
(complicated) proofs, and iii) suggest that the existence of $P_-$
may in fact be necessary (as well as sufficient) for the energy of a
designer gravity theory to be bounded below.
\end{abstract}

\maketitle

\section{Introduction}

It is well known that anti-de Sitter (AdS) gravity coupled to a
scalar field with mass at or slightly above the
Breitenlohner-Freedman bound \cite{BF} admits a large class of
boundary conditions, defined by an essentially arbitrary real
function $W$.  For all (regular) $W$, the conserved charges are well
defined and finite
\cite{Henneaux02,Henneaux04,HM2004,HH2004,Hertog2005,
Henneaux06,Amsel2006,Marolf2006}, despite the fact that the scalar
field falls off slower\footnote{We define $W$ such that $W=0$
corresponds to turning off the independent subleading term in the
asymptotic expansion of the scalar field.  The usual boundary
conditions instead keep this term and turn off the leading term.} than
usual. Theories of this type have been called designer gravity
theories \cite{HH2004}, because their dynamical properties depend
significantly on the choice of $W$  (see
e.g.~\cite{HM2004,HM2004b}).

In supergravity theories with a dual conformal field theory (CFT)
description, the AdS/CFT duality \cite{mald98, Aharony00} relates
$W$ to a potential term $\int W({\cal O}) \, dS$ in the CFT action, where
${\cal O}$ is the field theory operator that is dual to the bulk
scalar for $W=0$ boundary conditions \cite{witten, bss}. This led
\cite{HH2004} to conjecture that (a) there is a lower bound on the
gravitational energy in those designer gravity theories where $W$ is
bounded from below, and that (b) the solutions locally minimizing
the energy are given by the spherically symmetric, static soliton
configurations found in \cite{HH2004}.

More recently, the stability of designer gravity theories has been
studied using purely gravitational arguments. In particular, a
lower bound on the conserved energy in terms of the global minimum
of $W$ was rigorously proven within a specific AdS gravity theory by
relating the Hamiltonian charges to spinor
charges~\cite{Hertog2005}. Arguments were given in~\cite{Amsel2006}
suggesting that these bounds hold more generally.  However, it was
subsequently discovered~\cite{Hertog2006} that  solutions with
arbitrarily negative energy can be constructed numerically for
certain theories with $W \ge 0$. This raised a puzzle, which we
resolve in this paper.

Our resolution focuses on an auxiliary construct, the
``superpotential" $P(\phi)$ built from the bulk scalar potential
$V(\phi)$.  This superpotential is an important ingredient in
constructing the spinor charge.  In \cite{Amsel2006}, it was shown
that any $V(\phi)$ satisfying the Breitenlohner-Freedman bound {\it
perturbatively} admits an appropriate superpotential $P(\phi)$.  It
was of course recognized that the global existence of $P(\phi)$ was
required for the proof of an energy bound, and that this global
existence may impose constraints on $V(\phi)$.

What is interesting about the counter-examples of \cite{Hertog2006}
is that they {\it do} admit a globally defined superpotential, but
nevertheless violate the proposed bound.  The issue turns out to be
that superpotentials associated with a given $V(\phi)$ can be of two
types, which we call $P_+$ and $P_-$.   A particular $V(\phi)$ may
admit (distinct) superpotentials of both types, or it may admit only
the $P_+$ type.    The proof requires existence of a $P_-$-type
superpotential, while the examples of \cite{Hertog2006} admit only
the $P_+$-type\footnote{The fact that a given $V(\phi)$ can admit both $P_+$ and $P_-$ superpotentials was used to study asymptotically
AdS domain wall solutions in \cite{berg, FNSS, CCDVZ,Zag,ST1,ip,ST2, ST3}.}.

We verify this claim below and also confirm various other details of
the arguments of \cite{Hertog2005,Amsel2006}.  After briefly listing
our conventions in section \ref{prelim}, we review the proof of the
lower bound in section \ref{two}, illustrating why a $P_-$-type
superpotential is essential (and why a $P_+$-type superpotential is
not sufficient).  Section \ref{solitons} confirms that the
counter-examples of \cite{Hertog2006} admit {\it only} $P_+$-type
superpotentials and numerically explores the energy bound in a
number of examples.  We find evidence that the global existence of
$P_-$ may be necessary (as well as sufficient) for any lower bound
to hold.

Section \ref{spinors} investigates other aspects of the proof from
\cite{Hertog2005,Amsel2006}.  Arguments for positivity of the spinor
charge require Witten spinors, whose global existence can be
difficult to demonstrate. Questions about this global existence were
raised in \cite{Hertog2006}. However, we show that the argument for
global existence given in \cite{Hertog2005} for a particular $P_-$
extends to the general case.  As a check, we also explicitly
demonstrate the existence of Witten spinors in the context of
spherical symmetry by analyzing an associated ordinary differential
equation.  We evaluate these spinors numerically for a particular
example and use the results to check the relation between the spinor
and Hamiltonian charges derived in \cite{Amsel2006}.   We close with
some discussion in section \ref{disc}.

\section{Preliminaries}
\label{prelim}

Our conventions and our definition of asymptotically anti-de Sitter
spacetimes follow those of \cite{Marolf2005,Hertog2005,Amsel2006}.
In particular, we consider gravity theories minimally coupled to a
scalar field with Lagrangian density given by
\begin{equation}
\label{theory} {\bf L} = \frac{1}{2} \, d^d x \sqrt{ - g} \, [R -
(\nabla \phi)^2 - 2V(\phi)]  \, ,
\end{equation}
where we have set $8\pi G = 1$.  Here the scalar potential $V(\phi)$
is of the form
\begin{equation}
V(\phi) = \Lambda+\frac{1}{2} m^{2} \phi^{2} + \ldots.
\end{equation}
near $\phi=0$.   For simplicity we assume that $V$ is even. The
constant $\Lambda$ is the cosmological constant, given by
\begin{equation}
\Lambda = -\frac{(d-1)(d-2)}{2\ell^2},
\end{equation}
with $\ell$ a positive length that we may set to one by rescaling
the metric and scalar field. The spacetime dimension is denoted $d$,
and we assume that $d \ge 4$. We will furthermore assume that the
scalar field is tachyonic ($m^2 < 0$),  with mass in the
Breitenlohner-Freedman range~\cite{BF}
\begin{equation}
\label{range} m^2_{BF} \le m^2 < m^2_{BF} +1,
\end{equation}
where $m^2_{BF} =-(d-1)^2/4$.

The metric of exact AdS space of unit radius (and $\phi=0$), given
by
\begin{equation}
\label{pureads} ds^{2}_{0} = - \left(1 + r^2 \right)\, dt^2 +
\frac{dr^2}{1 + r^2} + r^2  d \omega^2_{d-2}\, ,
\end{equation}
is an exact solution of the theory, where $d\omega^2_{d-2}$ is the
unit-radius round metric on $S^{d-2}$. As part of our boundary
conditions, we assume that the metric of a general solution
asymptotically approaches (\ref{pureads}) in the manner
  described in \cite{Marolf2005,Hertog2005,Amsel2006}, and that
the scalar field is asymptotically of the form
\begin{equation}
\label{phi} \phi =  \frac{\alpha}{r^{\lambda_{-}}} + \frac{\beta}{
r^{\lambda_{+}}}  + \dots ,
\end{equation}
where
\begin{equation}
\label{roots} \lambda_{\pm} = {d-1 \pm \sqrt{(d-1)^2+ 4 m^2}\over 2}
\ .
\end{equation}

To obtain a well-defined dynamics for the linearized theory, it is
necessary to impose a boundary condition at $r=\infty$ on the scalar
field, i.e. we must impose a relation between $\alpha$ and $\beta$
in~\eqref{phi}. For example, one can impose $\alpha=0$, leaving
$\beta$ totally unspecified.  We refer to this option as ``fast
fall-off boundary conditions."  Alternatively, one may set $\beta=0$,
leaving $\alpha$ unspecified. One may also impose more general
boundary conditions of the form
\begin{equation}
\label{dW} \beta \equiv \frac{dW}{d\alpha} \, ,
\end{equation}
where $W(\alpha)$ is an arbitrary smooth function.   Under the AdS/CFT
duality, this function $W$ appears as an additional potential term in the action for the
dual field theory   \cite{witten, bss}.

\section{A tale of two superpotentials}
\label{two}

An elegant way to prove energy bounds is Witten's spinor
method~\cite{witten2}, which proceeds by constructing a manifestly
positive ``spinor charge,'' and then comparing it to the energy of
the gravitational solution. Witten's argument was originally
given in the context of asymptotically flat spacetimes, but it can
be generalized to the asymptotically AdS situation. When the matter
satisfies the dominant energy condition apart from the negative
cosmological constant term (regarded as a contribution
$-\frac{\Lambda}{8 \pi G} g_{ab}$ to $T_{ab}$), Witten's argument
may simply be modified by the addition  of a term
$\frac{1}{\sqrt{2(d-2)}} P \gamma_a$ to the covariant derivative
which acts on the spinor, where $P$ is a constant proportional to
$\sqrt{-\Lambda}$. This term is needed to deal with the negative
energy associated with the cosmological constant.

Our interest here is in tachyonic scalars $\phi$, whose potential
energy may in fact be unbounded below and does not satisfy the
dominant energy condition.  As shown by \cite{Boucher}
for $d=4$ (based on \cite{GHW}, and extended to higher dimensions by
\cite{town}), many such settings may be addressed by generalizing
the constant $P$ to a real ``superpotential'' $P(\phi)$ satisfying
 \begin{equation}
 \label{vtop} V(\phi) = (d-2) \left(\frac{dP}{d\phi}\right)^2 -(d-1)
 P^2 \, ,
\end{equation}
where $V(\phi)$ is the scalar potential.

Taking $\phi =0$ to be the AdS vacuum, we are interested in
potentials for which $V'(0)=0$.  The value $V(0)$ then determines
the cosmological constant.  With fast fall-off boundary conditions
($\alpha=0$), the proof \cite{Boucher,town} requires only that there
be a solution to (\ref{vtop}) with $P'(0)=0$ and $P(0) >0$.
Perturbatively, i.e., in the sense of formal power series,
a solution of this form always exists when $V''(0)$
sets the scalar field mass to satisfy the Breitenlohner-Freedman
bound.   In fact, there are two such perturbative solutions:
\begin{equation}
\label{series} P_{\pm}(\phi) = \sqrt{\frac{d-2}{2}} +
\frac{\lambda_\pm}{2\sqrt{2(d-2)}} \, \phi^2 + O(\phi^4) \, ,
\end{equation}
where $\lambda_\pm$ are given by (\ref{roots}). However, the proof
of \cite{Boucher,town} requires that $P(\phi)$ be well-defined (and
real) for all $\phi$, and this imposes further restrictions on
$V(\phi)$.

In both the fast \cite{Boucher,town} and slow fall-off cases
\cite{Hertog2005,Amsel2006}, the proof proceeds by using
$P_\pm(\phi)$ to construct a spinor charge $Q_\pm$.  In the fast
fall-off case one may show (e.g., following the basic method
outlined by \cite{davis}) that $Q_+ = E = Q_-$, where $E$ is the
conserved energy. The proof of \cite{Boucher,town} is identical no
matter which superpotential is used, and it is sufficient that only
$P_+$ exist\footnote{\label{exist}Both $P_+$ and $P_-$ exist near
$\phi=0$. However, a real solution ceases to exist if (\ref{vtop})
forces $P'(\phi)$ to become imaginary; i.e., if $V + (d-1)P^2 < 0$.
Since $P_+ > P_-$ near $\phi=0$, and since this in turn implies
$|P'_+|> |P'_-|$, one finds $|P_+|> |P_-|$ for all $\phi \neq 0$ and
global existence of $P_-$ implies global existence of $P_+$, but not
vice versa.}. However, in order to derive an energy bound with
slower fall-off conditions, \cite{Hertog2005,Amsel2006} assumed the
existence of $P_-$.  We verify in subsection \ref{critical} below
that this choice is critical in this context, and that $P_+$ alone
does not lead to an energy bound. This turns out to resolve the
issue raised in \cite{Hertog2006}.  A simple example is discussed in
section \ref{simple}.

\subsection{Choosing the right superpotential}
\label{critical}

We now quickly repeat the derivation of the energy bound from
\cite{Hertog2005,Amsel2006}, examining both the original argument
(using $P_-, Q_-$) and an analogous argument based on $P_+, Q_+$.
For either superpotential, the spinor charge is defined as
\begin{equation}
\label{charge} Q = \int_{C} \ast {\bf B} \,,
\end{equation}
where the integrand is the Hodge dual of a suitably defined Nester
two-form~\cite{n}
\begin{equation}
\label{Nester} B_{cd} = \bar{\Psi} \gamma_{[c} \gamma_d \gamma_{e]}
\widehat{\nabla}^e \Psi + \textrm{h.c.}\, ,
\end{equation}
and $C=\partial \Sigma$ is a surface at spatial infinity that bounds
a spacelike surface $\Sigma$. In (\ref{Nester}), $\Psi$ is a Dirac
spinor and the covariant derivative is defined in terms of $P$
($=P_\pm$) as
\begin{equation}\label{nablawidehatdef}
\widehat{\nabla}_a \Psi = \nabla_a \Psi + \frac{1}{\sqrt{2(d-2)}} \,
P(\phi) \gamma_a \Psi \, .
\end{equation}
We require that the spinor field $\Psi$ approaches a covariantly
constant spinor (i.e. a Killing spinor) of pure AdS, $\Psi_0$, at
infinity. We furthermore assume that asymptotically $-\bar \Psi
\gamma^a \Psi \to (\partial_t)^a$. Using Gauss's theorem we can
rewrite the spinor charge $Q$ ($=Q_\pm$) as
\begin{equation}
\label{Q2} Q = \int_{\Sigma} d (\ast {\bf B}) = \int_{\Sigma}
(\nabla^b B_{ab}) \, N^a dS\, ,
\end{equation}
where $dS$ is the integration element on $\Sigma$, and $N^a$ the
unit normal. Letting $i,j,\dots$ denote directions in the tangent
space of the surface $\Sigma$, one can then show \cite{town} that
the integrand of (\ref{Q2}) is
\begin{equation}
\label{divB} (\nabla^b B_{ab}) N^a = \left[ 2(\widehat{\nabla}_i
\Psi)^\dagger \widehat{\nabla}^i \Psi - 2 (\widehat{\nabla}_i
\Psi)^\dagger \gamma^i \gamma^j \widehat{\nabla}_j \Psi +
\lambda^\dagger \lambda \right] \, ,
\end{equation}
where
\begin{equation}
\lambda = \frac{1}{\sqrt{2}}\left(\gamma^a \nabla_a \phi
-\sqrt{2(d-2)} \, \frac{dP}{d\phi}\right) \Psi \,.
\end{equation}
The first and third terms in~(\ref{divB}) are manifestly nonnegative
as written.  A negative contribution from the second term can be
avoided by imposing the Witten condition \cite{witten2}
\begin{equation}
\label{witten} \gamma^i \widehat{\nabla}_i \Psi = 0 \,,
\end{equation}
which is essentially the spatial Dirac equation.  In
section~\ref{spinors}, we recall the argument~\cite{Hertog2005} that
globally smooth spinors satisfying~\eqref{witten} with the above
boundary conditions exist in designer gravity. This establishes $Q
\geq 0$.

However, an energy bound can be derived only once we relate $Q$ to
the physical energy $E$.  While these coincide for fast fall-off
boundary conditions, they differ in the slow fall-off case.  One
consequence of this is that $Q$ is not in general conserved.
Another is that $Q$ may now depend on whether the spinor charge is
defined using $P_+$ or $P_-$. That is, $Q_+ \neq Q_-$.

In~\cite{Hertog2005,Amsel2006},  the covariant phase space method
of~\cite{WL,wi,wz}, was used to show (following \cite{Marolf2005})
that energy in designer gravity takes the form\footnote{Expression
(\ref{generalE}) describes the generic case. Additional terms are
present for special cases. See \cite{Amsel2006}.} : \ben
\label{generalE} E =  - \int_C {\mathcal E}_{ab} \xi^a \, ds^b -
(\lambda_+ - \lambda_-)  \int_C \left[ W(\alpha)  -
\frac{\lambda_-}{d-1} \alpha \beta \right] \xi_b \, ds^b, \een where
$\xi =  \tfrac{\partial}{\partial t}$ is the time translation
conjugate to $E$, $ds^a$ is the integration element on the cut $C
\cong S^{d-2}$ of the AdS conformal boundary $\I$, and ${\mathcal
E}_{ab}$ is the suitably rescaled electric part of the Weyl tensor,
which is smooth at $\I$ as a consequence of our boundary conditions.
The numerical value of $E$ is independent of the particular choice
of the cut.

One can derive the relation between $Q_\pm$ and the energy
$E$~\cite{Hertog2005, Amsel2006} by expanding the metric and the
spinors in an asymptotic series in $1/r$, using the Witten equation,
and using Einstein's equation.  The result is:
\begin{equation}
\label{qew} Q_\pm = E -  \int_{C} \left[(\lambda_+ -
\lambda_-)W(\alpha)- (\lambda_{\pm} - \lambda_-)\alpha \beta \right]
d\omega + \frac{1}{2} \lim_{r \to \infty} (\lambda_{\pm}-\lambda_-)
\, r^{d-1-2\lambda_-} \int_{C_r} \alpha^2\, d\omega \,  ,
\end{equation}
where for simplicity we have chosen $C \cong S^{d-2}$ to be a cut of
constant $t$ so that  $d\omega$ is the integration element of the
unit sphere $S^{d-2}$. In the final term, $C_r$ is a large sphere of
radius $r$ in $\Sigma$.

Choosing $P_-$ and using  $Q_- \ge 0$ yields the bound
\begin{equation}
\label{bound} E \geq \textrm{Vol}(S^{d-2})  (\lambda_+ - \lambda_-)
\inf W \, ,
\end{equation}
where ${\rm Vol}(S^{d-2})$ is the volume of the unit-radius
$S^{d-2}$. On the other hand, for $\alpha \neq 0$ choosing $P_+$
causes the final term in (\ref{qew}) to diverge.  Since $E$ and $W$
are manifestly finite, it follows that $Q_+$ is infinite.  It is
also infinitely larger than $E$, and its positivity yields no lower
bound on $E$. Thus, the energy bound~(\ref{bound}) has been
established only if the theory admits a $P_-$ superpotential; the
existence of $P_+$ alone is not sufficient.

\subsection{A simple example}
\label{simple}

Let us consider a simple example shown in \cite{Hertog2006} to have
an energy $E$ which is unbounded below.  We take $d=4$ and the
superpotential
\begin{equation}
\label{simpleEx} P(\phi) = \left( 1+\frac{1}{2} \phi^2 \right)
\exp\left(-\frac{1}{16} \phi^4 \right) \, .
\end{equation}
This choice corresponds to $m^2 = -2$ (that is, $\lambda_- = 1,
\lambda_+ = 2$), and since the coefficient of $\phi^2$ is
$\frac{1}{2}$, we see from (\ref{series}) that (\ref{simpleEx})  is
of the $P_+$ type.  To see that the corresponding $P_-$ does not
exist, let us write~(\ref{vtop}) as
\begin{equation}
\label{dP} \frac{dP}{d\phi} = \frac{1}{\sqrt{2}} \sqrt{V + 3 P^2} \,
.
\end{equation}

\begin{figure}
\begin{center}
\includegraphics[width=3.5in]{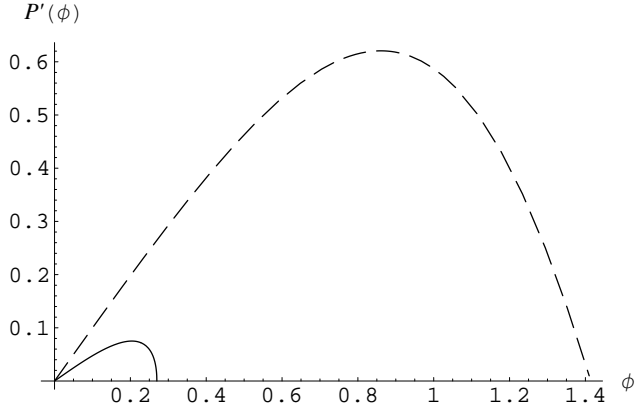}
\end{center}
\caption{Shown here are numerical plots of the derivative $P'(\phi)$
for the example presented in the text.  The dashed line corresponds
to the $P_+$ solution, which is stationary at the global minimum of
the potential, $\phi_{min} = \sqrt{2}$.  The solid line corresponds
to the $P_-$ solution, whose derivative vanishes at $\phi \approx
.27$. Hence, $P_-$ does not meet the global existence criterion. }
\label{Pprime}
\end{figure}

One may solve (\ref{dP}) by integrating out from $\phi = 0$ and
matching to the expansion (\ref{series}) for small $\phi$.  Such
solutions exist until the quantity $V + 3 P^2$ becomes negative.
Now, note that~(\ref{series}) implies $P_- < P_+$ for all nonzero
$\phi$ and similarly for their derivatives.  In our example, $P_+$
has a maximum at the global minimum of the potential occurring at
$\phi = \phi_{min}$, so $V+3P_+^2$ vanishes there.  But this means
that $V+3P_-^2$ must vanish at some $\phi_s < \phi_{min}$.  In fact,
it must vanish linearly at $\phi_s$, since $P_-'(\phi_s) =0$ would
force $V'(\phi_s)=0$ by (\ref{vtop}). Thus a real $P_-$ cannot exist
for $\phi > \phi_s$. This is illustrated in Figure~\ref{Pprime}
using numerical solutions of~(\ref{dP}).

Since only $P_+$ exists for all $\phi$, one does not expect
(\ref{bound}) to hold for solutions which explore large values of
$\phi$.  Indeed, \cite{Hertog2006} found solutions with $W \ge0$ and
arbitrarily negative energy
for this potential.

\section{Gravitational Solitons}
\label{solitons}

In order to confirm the above resolution of the puzzle raised in
\cite{Hertog2006}, we now numerically investigate energy bounds in a
simple class of examples.  For designer gravity boundary conditions
that preserve the full AdS symmetry group, the existence of solitons
has proven to be a reliable indicator whether or not a theory
satisfies a positive energy theorem
\cite{Heusler92,HH2004,Hertog2006,Hertog06b}. When the theory has a
static spherical soliton solution, no such theorem can hold because
AdS-invariant gravitational solitons can always be rescaled to
obtain solutions with arbitrarily negative energy that obey the same
boundary conditions\footnote{We emphasize that the soliton itself
has positive mass.}.

In this section we consider a one-parameter class
of scalar potentials and construct spherical solitons numerically.  We find that if the boundary conditions are specified by a $W$ which is i)
AdS-invariant and ii) bounded below, then solitons exist
{\it if and only if} $P_-$ does
not exist. This provides a strong test of the above stability proof,
and it suggests that the existence of $P_-$ is both a sufficient and
a necessary condition for the energy to be bounded from below in
designer gravity.

We consider the following class of potentials in $d=4$ dimensions:
\be\label{pot} V(\phi)=-3-  \phi^2  -{2 \over 3} \phi^6 +A \phi^8,
\ee
where $A>0$ is a free parameter. These yield scalar potentials
with a negative maximum at $\phi=0$, and with two global minima at
$\phi = \pm \phi_{min}$. Small fluctuations around $\phi=0$ have
$m^2 =-2$, which is within the range (\ref{range}). Hence
asymptotically the scalar generically decays as \be\label{genfall2}
\phi = {\alpha \over r }  + {\beta \over r^2} + \dots \ee and the
asymptotic behavior of the $g_{rr}$ metric component reads
\be\label{asmetric2} g_{rr} = {1 \over r^2} - { (1+\a^2/2 )\over
r^4} +{O} (r^{-5}). \ee

\begin{figure}
\begin{picture}(0,0)
\put(90,201){$\b$} \put(388,22){$\a$}
\end{picture}
\begin{center}
\includegraphics[width=3.5in]{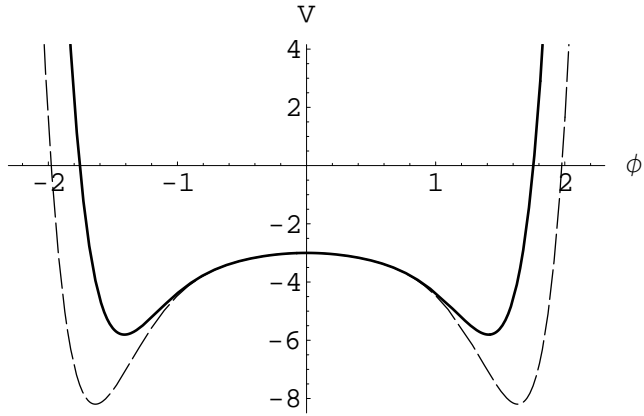}
\end{center}
\caption{The critical potential (solid line) that is on the verge of
violating the positive energy theorem for $W \geq 0$ designer
gravity boundary conditions. The dashed line shows a potential that violates
the positive energy theorem for designer gravity boundary conditions,
but satisfies such a theorem for fast fall-off boundary conditions ($\a=0$).}
\label{2}
\end{figure}

To construct the corresponding superpotentials $P_{\pm}$ we solve
(\ref{vtop}), starting with $P(0)=1$ and with $P''(0)={1 \over 4}
(3P(0) \pm \sqrt{9P(0)^2+4V''(0)})$. A solution to (\ref{vtop})
exists unless $V + 3 P^2$ becomes negative.  As we integrate out
from $\phi=0$, $P$ is initially increasing and $V + 3 P^2$ remains
positive because the scalar satisfies the Breitenlohner-Freedman
bound. For sufficiently large values of $A$ the global minima at
$\pm \phi_{min}$ will not be very much lower than the local maximum
at $\phi=0$, so one expects global solutions for $P_{\pm}$ to exist,
with $P'_{\pm}(\phi_{min})>0$. By contrast, if the global minima are
too deep, the quantity under the square root may become negative
before the global minimum is reached, and consequently a real
superpotential $P$ will not exist. Clearly there are two distinct
critical potentials (with parameters $A_c^\pm$), corresponding to
those where $V+3 P_{\pm}^2$ just vanishes as the global minimum is
reached. Numerically solving (\ref{vtop}) for a range of values for
$A$, one finds  $A_c^{+}=.1$, and $A_c^{-}=.283$. We plot the latter
critical potential in Figure 2 (full
curve). The dashed line in Figure 2 shows a potential with $A_c^{+}
\leq A < A_c^{-}$.

We now look for static spherical soliton solutions of the theory
(\ref{pot}) satisfying AdS-invariant boundary conditions with $W$ bounded below. Since $\lambda_+ =2$ and $\lambda_- = 1$,
AdS-invariant boundary conditions are given \cite{HM2004} by $W = k \alpha^3$ for real $k$.
We see that $W$ is bounded below only for $k=0$, so that  only solitons with $\beta=0$ lead to violations of (\ref{bound}).

The metric of a general
such solution takes the form \be \label{metric}
ds^2=-h(r)e^{-2\chi(r)}dt^2+h^{-1}(r)dr^2+r^2d\omega_2^2 , \ee and
the field equations read \be\label{hairy14d}
h\phi_{,rr}+\left(\frac{2h}{r}+\frac{r}{2}\phi_{,r}^2h+h_{,r}
\right)\phi_{,r}   =  V_{,\phi}, \ee \be\label{hairy24d}
1-h-rh_{,r}-\frac{r^2}{2}\phi_{,r}^2h =  r^2V(\phi), \ee \be
\label{hairy34d} \chi_{,r} = -{1 \over 2}r \phi_{,r}^2 . \ee
Regularity at the origin requires $h=1$ and
$h_{,r}=\phi_{,r}=\chi_{,r}=0$ at $r=0$. Rescaling $t$ shifts $\chi$
by a constant, so its value at the origin is arbitrary. Thus soliton
solutions can be labeled by the value of $\phi$ at the origin and
the set of all soliton solutions of a particular potential with a
negative maximum is found by integrating the field equations
(\ref{hairy14d})-(\ref{hairy34d}) for different values of $\phi$ at
the origin. For $ \vert \phi(0) \vert < \phi_{min}$ the scalar
asymptotically behaves as (\ref{genfall2}) with constant $\alpha,
\beta$.  The soliton therefore defines a point in the $(\a,\b)$
plane for each such $\phi(0)$. Repeating for all $\phi (0)$ yields a
curve $\b_{s}(\a)$. Given a choice of boundary condition $\b(\a)$,
the allowed solitons are simply given by the points where the
soliton curve intersects the boundary condition curve:
$\b_{s}(\a)=\b(\a)$.

Here we are interested in the existence of $\b=0$ solitons, for
potentials of the form (\ref{pot}).  Figure 3 shows how the value of
$\a$ for the $\b=0$ soliton changes when one increases the potential
parameter $A$, from $A \approx .2$ to its critical value
$A_c^{-}$. This corresponds to deforming $V$ from the potential
given by the dashed line in Figure 2, to the critical potential at
which $P_-$ begins to exist. One sees that $\a \rightarrow \infty$
precisely when $A \rightarrow A_c^{-}$. Furthermore, for $A >
A_c^{-}$ the $\b_{s}(\a)$ curve intersects the $\b=0$ axis only at
the origin, so no nontrivial
soliton solution exists in this parameter regime\footnote{When $A$ is further
 decreased to  $A_c^{+}$, an intersection point at finite $\b$
appears between $\b_s$ and the $\a=0$ axis.
This supports the claim of \cite{town} that the theory admits a positive energy theorem for $\a=0$ scalar boundary conditions only when
$V$ can be derived  from a superpotential ($P_{+}$).}. It appears,  therefore,
that regular spherical $W=0$ soliton solutions cease to exist precisely
when $A \rightarrow A_c^{-}$.

\begin{figure}
\begin{picture}(0,0)
\put(90,201){$A$} \put(388,22){$\a$}
\end{picture}
\begin{center}
\includegraphics[width=3.5in]{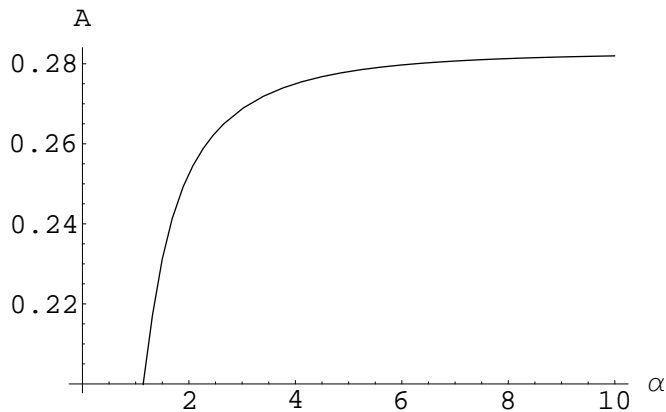}
\end{center}
\caption{ The value of $\a$ for the $W=0$ soliton in a range of
theories with different values for $A$, the coefficient of the
$\phi^8$ term in (\ref{pot}). One sees that regular spherical $W=0$
soliton solutions cease to exist precisely when $A \rightarrow
A_c^{-}=.283$.  } \label{3}
\end{figure}

The existence of scalar solitons with AdS-invariant boundary
conditions for $A < A_c^{-}$ implies there are negative energy
solutions in these theories. This can be seen as follows
\cite{Heusler92,Hertog04b}. Starting with a static soliton
$\phi_s(r)$, consider the one parameter family of configurations
$\phi_\l(r) = \phi_s(\l r)$. Because we chose conformally invariant
boundary conditions, these are again satisfied by the rescaled data.
It then follows from the constraint equations that the total energy of
the rescaled configurations takes the form \be E_\l = \l^{-3} E_1 +
\l^{-1} E_2 \ee where $E_2$ is independent of the potential and is
manifestly positive, and both $E_i$ are finite and independent of
$\l$. Furthermore, because the static soliton extremizes the energy
\cite{Sudarsky92}, one has
\be 0={d E_\l\over d\l}\Bigg|_{\l=1} = -3 E_1
-E_2
\ee
and hence $E_1=-{1 \over 3} E_2 <0$. Therefore the
contribution to the energy that scales as the volume, which includes
the potential and scalar terms, is negative. This means that
rescaled configurations\footnote{The rescaled configurations
$\phi_\l(r)$ are initial data for time dependent (but time-symmetric)
solutions. For
sufficiently small $\l$ one has a large central region where $\phi$
is essentially constant and away from an extremum of the potential.
Hence one expects the field to evolve to a spacelike `big crunch'
singularity \cite{Hertog04b}.}  $\phi_\l(r)$ with $\l < 1/\sqrt{3}$ must have
negative total energy, and hence violate the energy bound
(\ref{bound}).   Thus, the energy is unbounded below for potentials
with $A < A^-_c$ and $W=0$,  indicating that existence of $P_-$ is
necessary for energy in designer gravity
to be bounded below. (Note that we have already proven that it
is sufficient.)

\section{Existence of Witten Spinors}
\label{spinors}

As explained in section~\ref{two}, the derivation of the lower bound
for the energy $E$ uses the positivity of the spinor charge. This
requires the global existence of smooth spinors satisfying the
Witten condition \ben L \Psi = 0 \een on a constant time surface
$\Sigma$ with suitable asymptotic behavior, where $L$ is the
operator in the Witten equation,
\ben\label{witteneq}
L = \gamma^i \widehat \nabla_i
\, . \een
Here, $\widehat \nabla_i$ is the first order covariant 
differential operator 
defined in eq.~\eqref{nablawidehatdef}, taken in directions $i$ 
tangent to the spacelike surface $\Sigma$.

Global existence of such solutions is non-trivial to establish, and
indeed it was suggested in \cite{Hertog2006} that some subtle
failure of such spinors to exist might be responsible for the lack
of energy bounds in the systems studied there.  In contrast, we have
now proposed that the failure of those examples to satisfy
(\ref{bound}) was due solely to the fact that such potentials do not
admit a $P_-$-type superpotential.   In order to dispel any
remaining concerns about the global existence of spinors, we
reexamine this issue below.  Section \ref{again} essentially follows
\cite{Hertog2005} and proves global existence of the Witten spinors
in general, but we deviate slightly from~\cite{Hertog2005} to
correct an error in this proof. In order to correct this proof,
we need to work on a maximal slice $\Sigma$ (i.e., one for
which the trace of the extrinsic curvature satisfies
$K=0$) in the spacetime that intersects scri transversely.

This proof is in fact quite abstract and draws on
certain very non-trivial results in the mathematics literature. We
therefore restrict attention in section \ref{sphere} to the
particular case of maximal (e.g., time-symmetric), spherically
symmetric hypersurfaces, which is enough to address any concerns
raised by \cite{Hertog2006}. In this case we are able to give a more
explicit construction of the Witten spinors by reducing the problem
to ordinary differential equations.   Existence of solutions is then
straightforward to show. Finally, we numerically solve for the
Witten spinors in section \ref{numerics} and evaluate the
corresponding spinor charges, verifying the relationship between $Q$
and $E$ derived in section \ref{two}.

\subsection{The General Analytic Proof}
\label{again}
If $\Sigma$ is a maximal slice in the spacetime,
then the Witten operator $L$ in eq.~\eqref{witteneq}
can be written as
\ben\label{witteneq1}
L = \gamma^i \left( D_i + \frac{1}{\sqrt{2(d-2)}}P(\phi) \gamma_i
\right) \, ,
\een
where $D_i$ is now the spin-connection intrinsic to $\Sigma$.
On slices with $K \neq 0$, there would be another term proportional to
$K$ in the above expression for $L$. For the proof to work, we must assume that
such a slice exists. Below, we will briefly outline 
how one might construct such a $\Sigma$.

We would like to find a smooth
spinor field $\Psi$ on $\Sigma$ with the property that $L\Psi = 0$, and
that $\Psi$ tends to a suitable exact Killing spinor of AdS-space near infinity.
 Following \cite{Hertog2005}, the first step in the proof is to find
a formal power series solution $\eta$ to the Witten spinor equation
of the form
\ben
\label{asympt}
\eta \sim \Psi_0 + r^{-1} \Psi_1 +
r^{-2} \Psi_2 + \dots + r^{-N} \Psi_N + \dots \, , \een where
$\Psi_n \sim r^{1/2}$, and where $\Psi_0$ is a Killing spinor in
exact AdS-space (i.e., $\widehat \nabla_a \Psi_0 = 0$ in exact AdS
with vanishing $\phi$) such that \ben -\left(
\frac{\partial}{\partial t}\right)^a = \overline \Psi_0 \gamma^a
\Psi_0^{} \, . \een
The spinor fields $\Psi_1, \Psi_2, \dots$ are
determined recursively by the condition that $\eta$ be a formal
power series solution to the Witten equation. It can be proven that
$\Psi_n$ are uniquely determined for $n > 0$ by making a split of
the Witten equation into a part containing an $r$-derivative and a
part containing derivatives tangent to a sphere of constant $r$. The
explicit form of the first coefficients $\Psi_n$ in terms of the
spacetime curvature near infinity is given
in~\cite{Hertog2005, Amsel2006}. The formal power series solution
satisfies the equation \ben L\eta = J \,, \een where $J$ is a smooth
source vanishing faster than any inverse power of $r$ near $\I$. We
may terminate the expansion~\eqref{asympt} at some large finite $N$,
thus obtaining a $J$ vanishing faster than $r^{-N+1/2}$.

The next step is to obtain from the formal power series solution
$\eta$ a global solution $\Psi$ to the Witten equation. This step
requires some global analysis. The idea is to write the global
solution $\Psi$ that we seek  as \ben
 \Psi = \eta + \mu \, ,
 \een
where $\mu$ is a smooth spinor field to be determined.  Thus, $\mu$
should satisfy $L\mu = -J$, and $\mu$ should vanish at least as fast
as $r^{-1/2}$ in order that $\Psi$ have the same asymptotics as $\eta$
to leading order. The global existence of such a $\mu$ can be
established using the following key inequality~\eqref{ineq} which
holds whenever $(M,g_{ab}, \phi)$ is a solution to the Einstein
equations satisfying the asymptotic conditions given in
\cite{Hertog2005,Amsel2006}.   To state the inequality, consider any
smooth spinor field $u$ on $\Sigma$ that vanishes outside some compact
region in $\Sigma$, and introduce the following norm on such $u$:
 \ben
 \label{normu}
\|u \|^2 = \int_\Sigma \Bigg[(\widehat \nabla_i u)^\dagger \widehat
\nabla^i u + (1+r^2)^{-1} |u|^2 \Bigg]\, dS. \een
 Then there exists
a positive constant independent of $u$ such that \ben \label{ineq}
{\rm const.}^{-1} \| u \|^2 \le \int_\Sigma |L u|^2 \, dS \le {\rm
const.} \| u \|^2 \, . \een The proof of the inequality uses a
similar technique as in the ``Hardy-inequality,'' combined with the
key identity~\eqref{divB}. It may be found in~\cite{Hertog2005}.

We now use the inequality to establish a solution to the equation
$L\mu = -J$. We will do this by first constructing a distributional
solution $\mu$, and then showing that it is in fact smooth. Let $F$
be the linear functional on smooth compactly supported spinors $u$
defined by \ben F(u) = -\int_\Sigma ({\mathcal N}J)^\dagger u \, dS \, , \een
where ${\mathcal N} = \gamma_a N^a$, and $N^a$ is the unit timelike
normal to $\Sigma$. Note that ${\mathcal N}^2 = -1$, and that
\ben\label{hermitian}
{\mathcal N} L^\dagger = L {\mathcal N}
\een
due to the fact that $K=0$. By the Cauchy-Schwartz inequality and the first half of the
inequality~\eqref{ineq}, we get \bena |F(u)| &\le& \left(
\int_\Sigma |u|^2 (1+r^2)^{-1} \, dS \right)^{1/2} \left(
\int_\Sigma |J|^2 (1+r^2) \,
dS \right)^{1/2} \nonumber \\
&\le& {\rm const.} \left( \int_\Sigma |Lu|^2 \, dS \right)^{1/2} \,
. \eena The constant is finite because by construction $J$ drops off
sufficiently rapidly at infinity. We interpret this inequality as
saying that $F$ is a bounded functional with respect to the positive
definite scalar product given by \ben \langle v | u \rangle = \int
(Lv)^\dagger Lu \, dS \, . \een Let $H$ be the Hilbert space defined
by this inner product (which, by the second half of
inequality~\eqref{ineq} is identical to the Hilbert space obtained
from the norm $\|\, \cdot \, \|$). By the Riesz representation
theorem, there is hence an element $v \in H$ such that $F(u) =
\langle v | u \rangle$ for all $u \in C^\infty_0$. Again by the
inequality~\eqref{ineq}, every element in $H$ is locally square
integrable, so in particular a distribution on $\Sigma$. Thus, $v$
is a solution in the distributional sense of the equation $L^\dagger
L v = -{\mathcal N}J$. Next, we apply $\mathcal N$ to this equation,
use eq.~\eqref{hermitian}, and put $\mu = -{\mathcal N} L v$.
It follows that this $\mu$ is the desired distributional
solution to $L\mu = -J$, and $\Psi = \eta + \mu$ is a global
solution to the Witten equation.

It remains to prove that $\Psi$ is smooth, and that it satisfies the
desired boundary conditions. This will follow if we can show that
$\mu$ is smooth and vanishes sufficiently fast at infinity. It
follows from our construction so far only that $\mu \in L^2(\Sigma,
(1+r^2)^{-1}dS)$. But since $J$ is smooth and vanishes quickly, one can now use
the mapping properties of the parametrix of the operator $L$
established in~\cite{Mazzeo1989} to prove that $\mu$ is indeed
smooth and vanishes sufficiently fast. The details of this argument
are given in~\cite{Hertog2005}.

Finally, we outline how one might construct the desired $K=0$ slice $\Sigma$ in
the type of spacetimes that we consider. First, let $S$ be any (weakly) spacelike hypersurface
in spacetime, $D(S)$ its domain of dependence. If $S$ is such that the
closure of $D(S)$ is contained in a globally hyperbolic subregion of spacetime
(note that the full spacetime need {\em not} be globally hyperbolic), then a
theorem of Bartnik~\cite{Bartnik} guarantees that there exists a sufficiently
regular maximal slice $\Sigma$ with prescribed extrinsic curvature
and boundary $\partial \Sigma = \partial S$. Unfortunately, it is not immediately clear how to apply this theorem when the surface $S$ touches the conformal boundary.  However, by considering
a sequence $S_n$  eventually touching the conformal boundary transversely, we obtain a corresponding  sequence $\Sigma_n$ of maximal slices. Presumably it can be shown that these
have a subsequence that is convergent in a sufficiently strong sense.


\subsection{Spherical Symmetry}
\label{sphere}

We now explicitly describe the solution to the Witten equation for a slice
$\Sigma$ which is both spherically symmetric and maximal (i.e,
$K=0$) though not necessarily static. Such a slice automatically
exists when the initial data are time-symmetric.  Since this case is
sufficient to address the concerns of \cite{Hertog2006},  we explain
the analysis in some detail. Our conventions will be as follows. For
simplicity, we restrict attention in this subsection to the case
$d=4$.  We denote by $\hat{a}, \hat{b}, \ldots = 0,1,2,3$ indices on
a flat internal space (while $a,b, \ldots = t,r, \theta, \varphi$
are spacetime indices).   The gamma matrices satisfy
\begin{equation}
\gamma_{\hat{a}} \gamma_{\hat{b}} + \gamma_{\hat{b}}
\gamma_{\hat{a}} = 2 \eta_{\hat{a} \hat{b}} \,,
\end{equation}
and we choose the explicit matrix representation
\begin{equation}
\gamma^{\hat{0}} =  \left(
\begin{matrix}
0 & iI_2 \\
iI_2 & 0
\end{matrix}
\right) , \quad \gamma^{\hat j} = \left(
\begin{matrix}
0 & i\sigma^{\hat j}
\\ -i\sigma^{\hat j} & 0
\end{matrix}\right)
\end{equation}
where $I_2$ is the $2 \times 2$ identity matrix, and where
$\sigma^{1}, \sigma^2, \sigma^3$ are the standard Pauli matrices.

We wish to address spherical spacetimes. Let $\Sigma$ be a maximal
($K=0$), spherically symmetric spacelike slice on which we set
$t=0$, e.g., a surface of time-symmetry. Though such spacetimes are
not in general static, on $\Sigma$ they may nevertheless be written
in the form (\ref{metric}), which may be described by the
orthonormal frame
\begin{eqnarray}
e^{0} &=& \sqrt{h(r)}   e^{-\chi(r)} dt \\
e^{1} &=& r \sin\theta \,d\varphi \\
e^{2} &=& r d\theta \\
e^{3} &=& \frac{1}{\sqrt{h(r)}} \, dr \, .
\end{eqnarray}
Because $K=0$, the Witten equation~\eqref{witteneq} only depends on
the geometry intrinsic to $\Sigma$, i.e., $e^1, e^2, e^3$. We
further assume that the scalar field on $\Sigma$ is a function of
$r$ alone, and that $\phi(r)$, $h(r)$ are known functions. The
intrinsic covariant derivative acting on spinor fields is in general
given by
\begin{equation}
D_i \Psi = \partial_i \Psi + \frac{1}{4} \, \omega_i{}^{\hat{j}
\hat{k}} \gamma_{\hat j} \gamma_{\hat k} \Psi \,,
\end{equation}
where the intrinsic spin connection is determined by the relation
$-de^{\hat i} = \omega^{\hat i}{}_{\hat k} \wedge e^{\hat k} $. For
the metric~(\ref{metric}), we find
\begin{eqnarray}
\label{spin}
&&\omega^{1}{}_{ 2} = \cos \theta \,d\varphi  \\
&&\omega^{1}{}_{3} = \sqrt{h} \sin \theta \,d\varphi \\
&&\omega^{ 2}{}_{ 3} = \sqrt{h} \, d\theta \, .
\end{eqnarray}

With the above ingredients in hand, the Witten condition becomes the
explicit equation
\begin{equation}
\label{witcon} 0 = \gamma_{3} \sqrt{h} \left(\partial_r +
\frac{1}{r} \right) \Psi + \frac{1}{r} \left[\gamma_{2}
\left(\partial_{\theta}
 + \frac{\cos \theta}{2  \sin\theta}\right)   + \frac{1}{\sin\theta}\gamma_{1} \partial_{\varphi}  \right]\Psi
 +\frac{3}{2} P(\phi) \Psi \,.
 \end{equation}
Writing $\Psi$ in terms of two-component spinors as $\Psi =
\left(\begin{array}{c} \Psi_1 \\ \Psi_2 \end{array} \right) \,,$
(\ref{witcon}) implies
\begin{eqnarray}
\label{witcon21} 0 &=& i \left[\sigma_{3} \sqrt{h} \left(\partial_r
+ \frac{1}{r} \right) + \frac{1}{r} \, \rlap/\!\nabla_{\! \!S^2}
\right]\Psi_2
 + \frac{3}{2} P(\phi) \Psi_1, \\
 \label{witcon22}
0 &=& i \left[\sigma_{3} \sqrt{h} \left(\partial_r + \frac{1}{r}
\right) + \frac{1}{r} \, \rlap/\!\nabla_{\! \!S^2} \right]\Psi_1
 - \frac{3}{2} P(\phi) \Psi_2 \,.
\end{eqnarray}
Here the operator $\rlap/\!\nabla_{\! \!S^2}$ is the Dirac operator
on the 2-dimensional sphere $S^2$.
To decouple the differential equations~(\ref{witcon21})
and~(\ref{witcon22}), we define
 \begin{equation}
 \Psi_+ \equiv \frac{1}{2} (\Psi_1 + i \Psi_2), \quad
 \Psi_- \equiv \frac{1}{2}(\Psi_1 - i \Psi_2) \, .
 \end{equation}
Then the $\Psi_\pm$ satisfy
\begin{eqnarray}
\label{witcon31} \left[ \sigma_{3} \left(\partial_r + \frac{1}{r}
\right) + \frac{1}{r \sqrt{h}} \, \rlap/\!\nabla_{\! \!S^2} +
\frac{3}{2 \sqrt{h}} P(\phi) \right] \Psi_+ &=& 0 \\
\label{witcon32} \left[ \sigma_{3} \left(\partial_r + \frac{1}{r}
\right) + \frac{1}{r \sqrt{h}} \, \rlap/\!\nabla_{\! \!S^2} -
\frac{3}{2 \sqrt{h}} P(\phi) \right] \Psi_- &=& 0 \, .
\end{eqnarray}

Next we wish to expand $\Psi_\pm$ in spinor spherical harmonics,
which are eigenfunctions of $\rlap/\!\nabla_{\! \!S^2}$.  These are
given, for example in~\cite{ch}, by
\begin{equation}
\Theta^{(-)}_{\pm n l}(\theta, \varphi) = \left(\begin{array}{c}
\left(\cos \frac{\theta}{2}\right)^{l+1}
 \left(\sin \frac{\theta}{2}\right)^l P^{(l, \,l+1)}_{n-l}(\cos \theta)  \\
  \pm i \left(\cos \frac{\theta}{2}\right)^{l}
\left(\sin \frac{\theta}{2}\right)^{l+1} P^{(l+1, \,l)}_{n-l}(\cos
\theta)
 \end{array} \right) e^{-i (l+\frac{1}{2})\varphi}
 \end{equation}
 \begin{equation}
\Theta^{(+)}_{\pm n l}(\theta, \varphi) = \left(\begin{array}{c} i
\left(\cos \frac{\theta}{2}\right)^{l}
 \left(\sin \frac{\theta}{2}\right)^{l+1} P^{(l+1, \, l)}_{n-l}(\cos \theta)  \\
  \pm \left(\cos \frac{\theta}{2}\right)^{l+1}
\left(\sin \frac{\theta}{2}\right)^{l} P^{(l,\, l+1)}_{n-l}(\cos
\theta)
 \end{array} \right) e^{i (l+\frac{1}{2})\varphi} \, ,
 \end{equation}
where $n,l$ are integers such that $n, l \geq 0$ and $n \geq l$.
The $P^{(a, b)}_n(x)$ are Jacobi polynomials and the spherical
harmonics satisfy
\begin{equation}
\rlap/\!\nabla_{\! \!S^2} \Theta^{(s)}_{\pm n l} = \pm i (n+1)
\Theta^{(s)}_{\pm n l} \,.
\end{equation}
Separating variables as
\begin{equation}
\label{ansatz} \Psi_+(r, \theta, \varphi) = \sum_{s, \pm, n, l}
R^{(s)}_{\pm n l} (r) \Theta^{(s)}_{\pm n l} (\theta, \varphi)
\end{equation}
\begin{equation}
\label{tansatz} \Psi_-(r, \theta, \varphi) = \sum_{s, \pm, n, l}
\tilde{R}^{(s)}_{\pm n l} (r) \Theta^{(s)}_{\pm n l} (\theta,
\varphi) \,,
\end{equation}
we find from~(\ref{witcon31}) and~(\ref{witcon32}) that the radial
functions satisfy the following differential equations:
\begin{equation}
\label{radialde} \frac{d R^{(s)}_{\mp n l} }{dr}+ \frac{1}{r}
R^{(s)}_{\mp n l}+ \frac{3}{2 \sqrt{h}} P(\phi) R^{(s)}_{\pm n l}
\pm i(n+1) \frac{1}{r \sqrt{h}} R^{(s)}_{\pm n l} = 0
\end{equation}
\begin{equation}
\label{tradialde} \frac{d \tilde{R}^{(s)}_{\mp n l} }{dr}+
\frac{1}{r} \tilde{R}^{(s)}_{\mp n l}- \frac{3}{2 \sqrt{h}}P(\phi)
\tilde{R}^{(s)}_{\pm n l}
 \pm i(n+1) \frac{1}{r \sqrt{h}} \tilde{R}^{(s)}_{\pm n l} = 0 \,.
\end{equation}

To pick the desired solution to the Witten equation, we now impose
our boundary conditions. We demand that our solution be regular in
the interior, and that it asymptotically approaches a suitable
Killing spinor $\Psi_0$ of pure AdS.  As we will see shortly, this
$\Psi_0$ has only $n=l=0$ components. Since the remaining modes are
decoupled, they may be consistently set to zero\footnote{In fact, a
careful analysis shows that the boundary conditions force such modes
to vanish.}.  Thus we consider only $n= l=0$ below.

Near the origin $r=0$, the solutions  behave as
\begin{equation}
\label{near0} R^{(s)}_{- 0 0 } \sim \frac{i c_1}{r^2} - i c_2 \,,
\quad R^{(s)}_{+ 0 0 } \sim \frac{c_1}{r^2} + c_2
\end{equation}
where $c_1, c_2$ are constants.  Regularity at the origin then
requires boundary conditions of the form
\begin{equation}
\label{wcbc} R^{(s)}_{- 0 0 }(0) = -i c, \quad R^{(s)}_{+ 0 0 }(0) =
c \, .
\end{equation}
The analysis for the $\tilde{R}^{(s)}_{\pm 0 0 }$ solutions near the
origin is exactly the same. As $r \to \infty$, we require that the
solution to the Witten condition approaches a Killing spinor in
exact AdS space. For large $r$, solutions to~(\ref{radialde}) and~(\ref{tradialde}) behave
as
\begin{eqnarray}
\label{largerR}
R^{(s)}_{- 0 0 } &\sim& \frac{c_3}{r^{5/2}} - c_4
r^{1/2} \,, \quad \quad R^{(s)}_{+ 0 0 } \sim \frac{c_3}{r^{5/2}} + c_4
r^{1/2} \nonumber \\
\tilde R^{(s)}_{- 0 0 } &\sim& -\frac{\tilde c_3}{r^{5/2}} + \tilde c_4
r^{1/2} \,, \quad \tilde R^{(s)}_{+ 0 0 } \sim \frac{\tilde c_3}{r^{5/2}} + \tilde c_4
r^{1/2} \,.
\end{eqnarray}
 The solution which grows as $r^{1/2}$ matches the known
asymptotic behavior of Killing spinors in exact AdS space.  In
particular, an explicit solution to the Killing spinor equation is
given by \cite{Henneaux85}
\begin{equation}
\label{ks} \Psi_0( t = 0, r, \theta, \varphi)  =
\left(\rho_1(r)-\rho_2(r) \gamma_{3}\right) \left(\cos
\frac{\theta}{2}+ \sin \frac{\theta}{2} \gamma_{3} \gamma_{2}\right)
\left(\cos \frac{\varphi}{2}+ \sin \frac{\varphi}{2} \gamma_{2}
\gamma_{1}\right) U\,,
\end{equation}
where $U$ is a constant spinor and
\begin{equation}
\rho_1(r) = \left(\frac{(1+r^2)^{1/2}+1}{2}\right)^{1/2}\,, \quad
\rho_2(r) = \left(\frac{(1+r^2)^{1/2}-1}{2}\right)^{1/2} \,.
\end{equation}
The spinor $U$ is chosen so that $\Psi_0$ satisfies the additional
requirement $(\partial_t)^a = -\bar \Psi_0 \gamma^a \Psi_0$, or
equivalently $\Psi^{\dagger}_0 \Psi_0 \sim r$.  One can achieve this
normalization by taking
\begin{equation}
\label{norm} U =  \frac{1+i}{\sqrt{2}} \left(\begin{array}{c} 1 \\ 0
\\ 0 \\0 \end{array} \right) \,.
\end{equation}
With the choice~(\ref{norm}), one can rewrite~(\ref{ks}) in terms of
the spinor spherical harmonics as
\begin{equation}
\label{ks2}
\Psi_0 =  \left(\begin{array}{c} \frac{\rho_1}{\sqrt{2}}\left(\Theta^{(-)}_{- 0 0}+ i \Theta^{(-)}_{+ 0 0}\right) \\
\frac{i \rho_2}{\sqrt{2}}\left(i \Theta^{(-)}_{- 0 0}+
\Theta^{(-)}_{+ 0 0}\right) \end{array} \right) \,.
\end{equation}
Hence, the solution for the Witten spinor is
\begin{equation}
\label{witspin} \Psi = \left(\begin{array}{c} \left(R^{(-)}_{- 00}
+\tilde{R}^{(-)}_{- 00} \right) \Theta^{(-)}_{- 0 0} +
\left(R^{(-)}_{+ 00} +\tilde{R}^{(-)}_{+ 00} \right) \Theta^{(-)}_{+
0 0} \\ -i\left[\left(R^{(-)}_{- 00} -\tilde{R}^{(-)}_{- 00} \right)
\Theta^{(-)}_{- 0 0}+  \left(R^{(-)}_{+ 00} -\tilde{R}^{(-)}_{+ 00}
\right) \Theta^{(-)}_{+ 0 0} \right] \end{array} \right) \,.
\end{equation}
Furthermore, comparing to~(\ref{ks2}) we see that the asymptotic
conditions on the radial functions are
\begin{eqnarray}
\label{wcasym} R^{(-)}_{- 0 0} \to \frac{1}{4}  (1-i)
r^{1/2}, \quad R^{(-)}_{+ 0 0} \to \frac{1}{4} (i-1) r^{1/2}
\nonumber \\
\tilde{R}^{(-)}_{- 0 0} \to \frac{1}{4} (1+i) r^{1/2}, \quad
\tilde{R}^{(-)}_{+ 0 0} \to \frac{1}{4} (1+i) r^{1/2} .
\end{eqnarray}

Comparing (\ref{wcasym}) and (\ref{wcbc}) with (\ref{near0}) and
(\ref{largerR}) demonstrates that the desired Witten spinors exist.
One first notes that $R^{(s)}_{\pm00}$ decouple from  $\tilde
R^{(s)}_{\pm00}$.  To construct the desired $R^{(s)}_{\pm00}$, one
simply enforces the boundary condition (\ref{wcbc}) at $r=0$ for
some $c$ and integrates (\ref{radialde}) outward.  From
(\ref{largerR}), any solution will grow asymptotically as $r^{1/2}$,
and in fact will satisfy (\ref{wcasym}) up to an overall scale.
Rescaling $c$ yields the desired solutions $R^{(s)}_{\pm00}$.  The
analysis for $\tilde R^{(s)}_{\pm00}$ is similar.  Thus, Witten
spinors exist for any spherically symmetric maximal hypersurface
$\Sigma$ satisfying our boundary conditions.

\subsection{Numerical results}
\label{numerics}

A final crucial aspect of the argument of section \ref{two} is the
detailed relation (\ref{qew}) between the spinor charges $Q_\pm$ and
the energy $E$.  Since (\ref{qew}) was derived via a tedious
calculation, it is useful to verify the relation numerically.   Let
us consider the theory with potential
\begin{equation}
\label{veg} V(\phi) = -3- \phi^2 - \frac{2}{3} \phi^6 + \frac{1}{2}
\phi^8 \,,
\end{equation}
which is just (\ref{pot}) for $A=1/2 > A_c^-$.  This theory admits
both a $P_+$ and a $P_-$ superpotential.  We take as initial data a
soliton with $\phi(0) = 1/2$ and boundary condition $W(\alpha) = k
\alpha$ for some constant $k\neq 0$ (so that $W$ is not conformally
invariant).  We find that a soliton exists with $\beta_s = k \approx
-.566 $ and $\alpha_s \approx .787$.  The energy of this soliton is
$E \approx -2.956$.

\begin{figure}
\begin{center}
\includegraphics[width=3.5in]{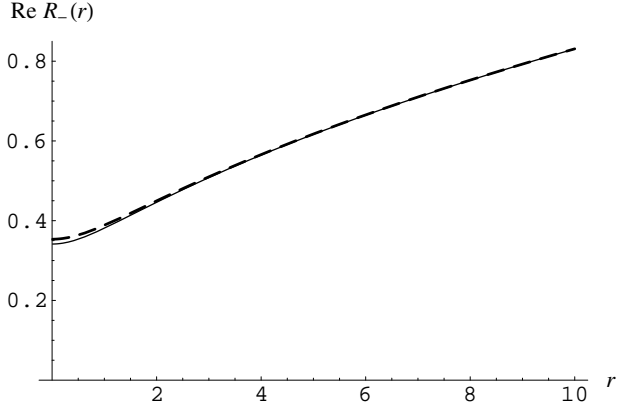}
\end{center}
\caption{Numerical results for the example~(\ref{veg}) confirm the
existence of Witten spinors that are regular at the origin and have
the correct behavior at large $r$. The solid line is $\textrm{Re}\,
R_- = \textrm{Im}\, R_+ = \textrm{Re}\, \tilde R_- =
\textrm{Im}\,\tilde R_+$, which by the arguments in the text
asymptotically approach $\frac{\rho_1}{2\sqrt{2}}$, shown here as
the dashed line.} \label{RRm}
\end{figure}

\begin{figure}
\begin{center}
\includegraphics[width=3.5in]{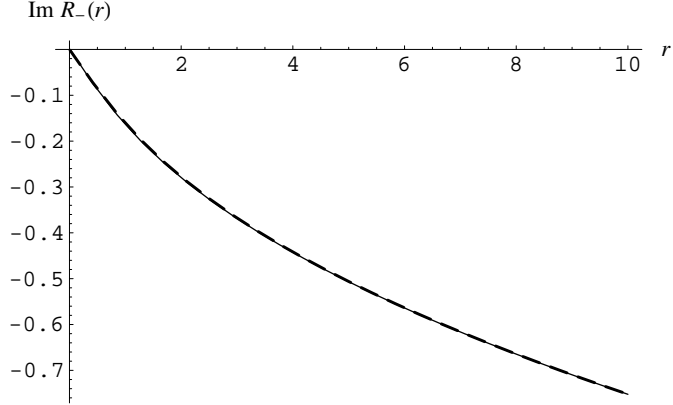}
\end{center}
\caption{The solid line is $\textrm{Im}\, R_- = \textrm{Re}\, R_+ =
-\textrm{Im}\, \tilde R_- = -\textrm{Re}\,\tilde R_+$, which by the
arguments in the text asymptotically approach
$-\frac{\rho_2}{2\sqrt{2}}$, shown here as the dashed line. }
\label{IRm}
\end{figure}

Numerical results for $P_-$-type Witten spinors are shown in
Figures~\ref{RRm} and~\ref{IRm}.\footnote{In theories containing AdS
invariant solitons, we have also verified numerically the existence
of the Witten spinors (for $P_+$)  with initial data given by the
rescaled configurations $\phi_{\lambda} (r) = \phi( \lambda r)$
discussed in section~\ref{solitons}.}
Using these solutions, we can also calculate the spinor charge and
check~(\ref{qew}).  We obtain $Q_- \approx 2.655$, thus confirming
the relation $E = Q_- + 4\pi W$. Similarly, for the
$P_+$ case we find $(Q_+ - E + 4\pi(W- \alpha \beta))/(2 \pi
\alpha^2 r) \approx 1.00$, in agreement with (\ref{qew}).

\section{Discussion}
\label{disc}

We have resolved the puzzle raised in \cite{Hertog2006} concerning
energy bounds in designer gravity.  While the arguments of
\cite{Hertog2005,Amsel2006} are correct as written, global existence
of an appropriate superpotential is a subtle requirement. In
particular, two types of superpotentials may (or may not) exist for
a given scalar potential $V(\phi)$.  The proof of
\cite{Hertog2005,Amsel2006} requires the global existence of a
$P_-$-type superpotential.  If one attempts to follow the same
argument with a $P_+$-type superpotential, one finds that the
difference between the conserved energy $E$ and the associated
spinor charge $Q_+$ diverges, and in particular that $Q_+$ diverges.
Thus, positivity of $Q_+$ does not yield a lower bound for $E$, and
the existence of a $P_+$-type superpotential alone is not sufficient
to yield a positive energy theorem.  Numerical explorations support
this resolution.

A specific question raised in \cite{Hertog2006} concerned the global
existence of Witten spinors. We have demonstrated (section
\ref{spinors})  that no such difficulties arise, even for the models
considered in \cite{Hertog2006}.   The existence theorem of
\cite{Hertog2005} was shown to hold in general and, for the special
case of spherical symmetry, existence of Witten spinors on maximal
(e.g., time-symmetric) hypersurfaces was again demonstrated using
simple arguments based on ordinary differential equations. The
spherical time-symmetric context is sufficient to address the
concerns of \cite{Hertog2006}.  This reinforces our claim that, when
$W$ has a global minimum,  the existence of a $P_-$-type
superpotential is sufficient for the energy to be bounded below.

Interestingly, our numerical explorations of particular examples
(section \ref{solitons}) suggest that global existence of $P_-$ may
also be {\it necessary}  for the energy to be bounded below.
Such a property would be
analogous to the claim of \cite{town} that the existence of a superpotential is
necessary for an energy bound to hold in the case of fast fall-off boundary conditions $(\alpha=0)$.
However, in the fast fall-off case the relevant superpotential is $P_+$, since either superpotential is sufficient
and existence of $P_-$ implies existence of $P_+$ (footnote \ref{exist}).    Indeed, for $\alpha=0$ the spinor charges $Q_\pm$
satisfy $Q_\pm = E$ and, in particular, $Q_+=Q_-$ when both potentials exist.   Thus we find that each type of superpotential provides a
stability criterion for AdS gravity-scalar theories, with $P_+$ controlling the $\alpha=0$ case and $P_-$ controlling cases with slower
fall-off conditions on the scalar.

Finally, a key tool in our investigation of particular examples was
the argument from \cite{Heusler92,HH2004,Hertog2006,Hertog06b}
showing that  the existence of designer gravity solitons satisfying
AdS-invariant boundary conditions implies that the energy is
unbounded below.    We may thus interpret the lower bounds of
\cite{Hertog2005,Amsel2006} in terms of the soliton content of such
theories.  We conclude that
designer gravity theories where $P_-$ exists
can have no solitons when $W$ is both AdS-invariant and bounded below.

\begin{acknowledgments}
The authors would like to thank Gary Horowitz for useful
discussions. S.H. would also like to thank Piotr Chrusciel and Bob Wald for
discussions and the physics department at UCSB for its
hospitality. D.M. also thanks Jim Isenberg and Robert Bartnik for discussions.
A.A. and D.M. were supported in part by NSF grant
PHY0354978, and by funds from the University of California.
\end{acknowledgments}

\end{document}